\documentclass[a4paper,12pt]{article}

\usepackage[margin=2cm]{geometry}
\usepackage{amsmath}
\usepackage{graphicx}
\DeclareGraphicsExtensions{.pdf,.png,.PNG,.jpg}
\usepackage{cite}

\begin{document}
% Title portion. Note the short title for running heads 
\title{\large{\textbf{Deep Neural Network Optimized to Resistive Memory with Nonlinear Current-Voltage Characteristics}}}
\author{\normalsize{Hyungjun Kim$^{\ast}$, Taesu Kim\thanks{Hyungjun Kim and Taesu Kim eqaully contributed to this work.}, Jinseok Kim, and Jae-Joon Kim\thanks{To whom correspondence should be addressed; E-mail: jaejoon@postech.ac.kr}}}
\date{%
    \small{Department of Creative IT Engineering, POSTECH, Pohang, South Korea}
}
\maketitle
\begin{abstract}
Artificial Neural Network computation relies on intensive vector-matrix multiplications. Recently, the emerging nonvolatile memory (NVM) crossbar array showed a feasibility of implementing such operations with high energy efficiency, thus there are many works on efficiently utilizing emerging NVM crossbar array as analog vector-matrix multiplier. However, its nonlinear I-V characteristics restrain critical design parameters, such as the read voltage and weight range, resulting in substantial accuracy loss. In this paper, instead of optimizing hardware parameters to a given neural network, we propose a methodology of reconstructing a neural network itself optimized to resistive memory crossbar arrays. To verify the validity of the proposed method, we simulated various neural network with MNIST and CIFAR-10 dataset using two different specific Resistive Random Access Memory (RRAM) model. Simulation results show that our proposed neural network produces significantly higher inference accuracies than conventional neural network when the synapse devices have nonlinear I-V characteristics.
\end{abstract}

\section{Introduction}

In recent years, Artificial Neural Network (ANN) has been gaining significant interest by claiming several cutting-edge results in solving various nonlinear problems \cite{bengio}. The breakthrough of ANN heavily depends on the expansion of networks in depth, which requires vast amount of vector-matrix multiplications. With the advent of vector-matrix multiplication acceleration based on graphics processing units (GPUs), large and deep neural networks have been able to handle complex tasks using extensive amounts of data \cite{GPU}. However, despite the fact that GPUs provide highly parallel computing suitable for ANNs, the high power consumption of GPUs is an obstacle to be improved. To address the issue, many dedicated accelerators for vector-matrix multiplications have been proposed \cite{RPU,DPE,ISAAC,PRIME}.

\begin{figure}[t]
  \includegraphics[width=0.8\textwidth, height=0.44\textwidth]{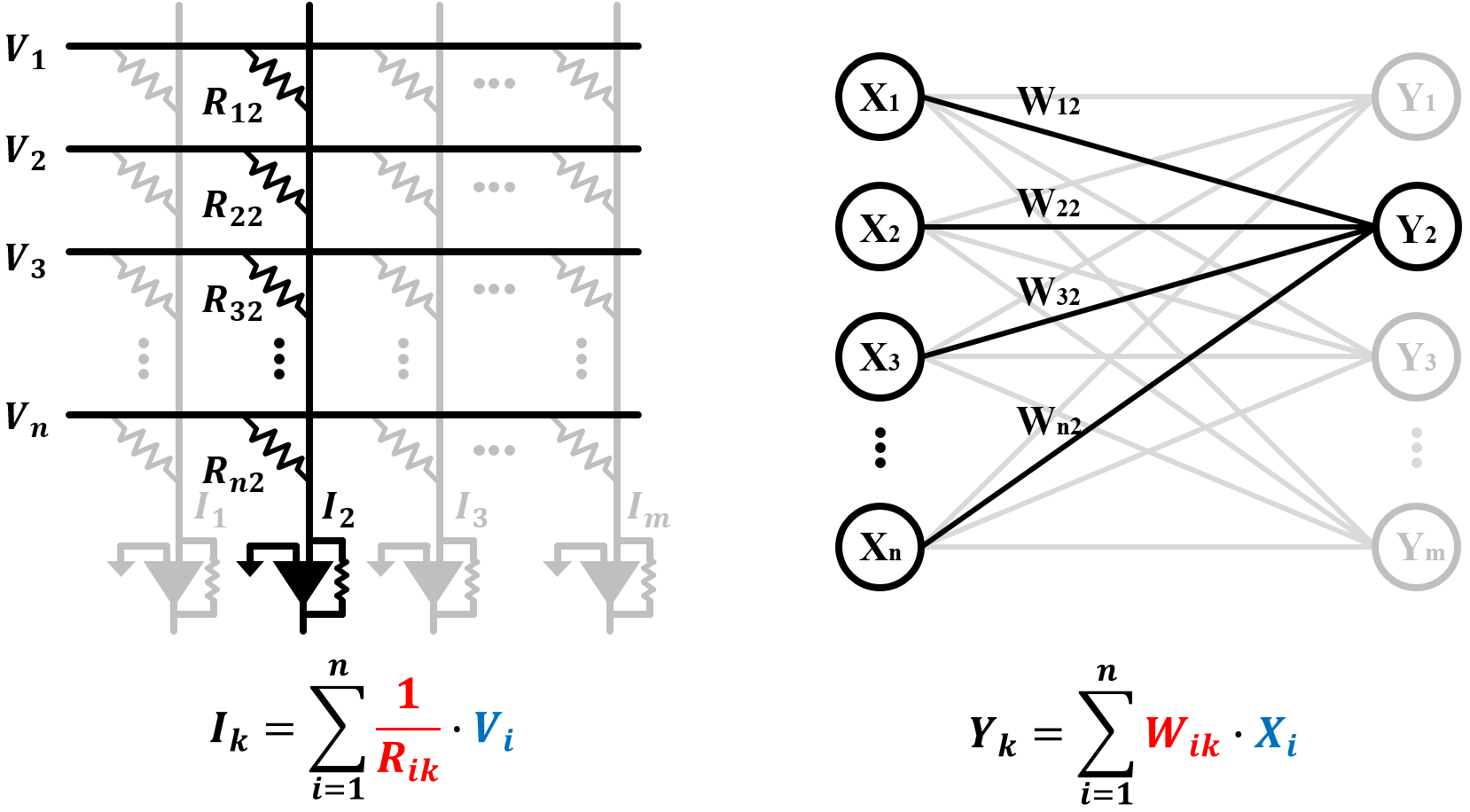}
  \centering
  \caption{\small An example of mapping vector-matrix multiplication to a RRAM crossbar array}
  \label{crossbar}
\end{figure}

Emerging nonvolatile memory (NVM) technologies including Phase-Change Random Access Memory (PCRAM), Resistive Random Access Memory (RRAM), Conductive-Bridge Random Access Memory (CBRAM), and Spin-Transfer-Torque Magnetic Random Access Memory (STT-MRAM) have been widely studied as next generation memories \cite{EMT}. While conventional memories such as Static Random Access Memory (SRAM), and FLASH memory are charge-based, emerging NVM is current-based and represents states with different resistance values.
This current-based nature opens up the opportunity to use emerging NVM for neural network acceleration. Current-based devices in a crossbar array structure can straightforwardly implement vector-matrix multiplication in neural network computations as shown in Fig. \ref{crossbar}. By mapping input vector to input voltages and weight matrix to resistive crossbar array, vector-matrix multiplication can be calculated in a single step by sampling the current flowing in each column \cite{MH}. Since this approach can be several orders of magnitude more efficient than CMOS ASIC approaches in terms of both speed and power \cite{DPE,RPU,PRIME,ISAAC}, many studies proposed neural network accelerators based on emerging NVM crossbar array \cite{nature,SEI,BurrReview,Burr}.

However, there are several issues with using emerging NVM crossbar array as an analog multiplier. Sneak path problem is one of the most critical issues \cite{EMT,BurrReview,nature2,sneak_path}. Various works have tried to solve this problem in different ways \cite{sneak_path,selector,EMT}. The most common idea is to use a device with nonlinear I-V characteristics. For example, it has been proposed to serially connect a selector device such as a transistor or a diode to an emerging NVM cell or to make the I-V characteristic of an emerging NVM cell as nonlinear as possible \cite{EMT}. However, although this approach can overcome the sneak path problem, it degrades the accuracy of current-based vector-matrix multiplication because nonlinear I-V characteristics hinder precise implementation of linear multiplications required for vector-matrix multiplications. Several works tried to solve this issue by restricting the range of the reading voltage to use the pseudo-linear sub-region of the nonlinear I-V curve \cite{penggu,approximate}. However, limiting reading voltage worsened DAC resolution issue, making it difficult to compute complex neural networks using emerging NVM crossbar arrays. Another previous approach was to address the problem by tuning the weights considering the computational error before mapping \cite{DPE,DPE2}. However, it also failed to utilize full input voltage range. In addition, these approaches did not fully address how the increase of the nonlinearity of I-V characteristics affects the inference accuracy of neural networks.

Unlike previous approaches which attempted to precisely map pre-determined weights to the crossbar array to reduce accuracy loss, we propose to rather construct an ANN model itself which accomodates nonlinear I-V characteristics. This allows nonlinear I-V characteristics to be taken into account during both of the learning phase and the inference phase of a neural network, reducing discrepancies between neural network models and emerging NVM-based hardwares. In this paper, we have selected two nonlinear RRAM devices as proof-of-concept devices and performed simulations to verify the idea based on the characteristics of the devices. The main contribution of this paper is as follows:

1) We analyze the correlation between the degree of nonlinearity of I-V characteristics and the inference accuracy loss in RRAM-based ANN hardware. We show that the degree of accuracy loss depends on the distribution of activation values.

2) We propose a modified perceptron model that is compatible with the nonlinear I-V characteristics of resistive memory devices and demonstrate how to train neural networks based on the proposed model. We show that neural networks based on the proposed model can avoid the loss of inference accuracy which happens while mapping the network to RRAM crossbar arrays.

\section{Preliminaries}
\begin{figure}[t]
  \includegraphics[width=\textwidth, height=0.38\textwidth]{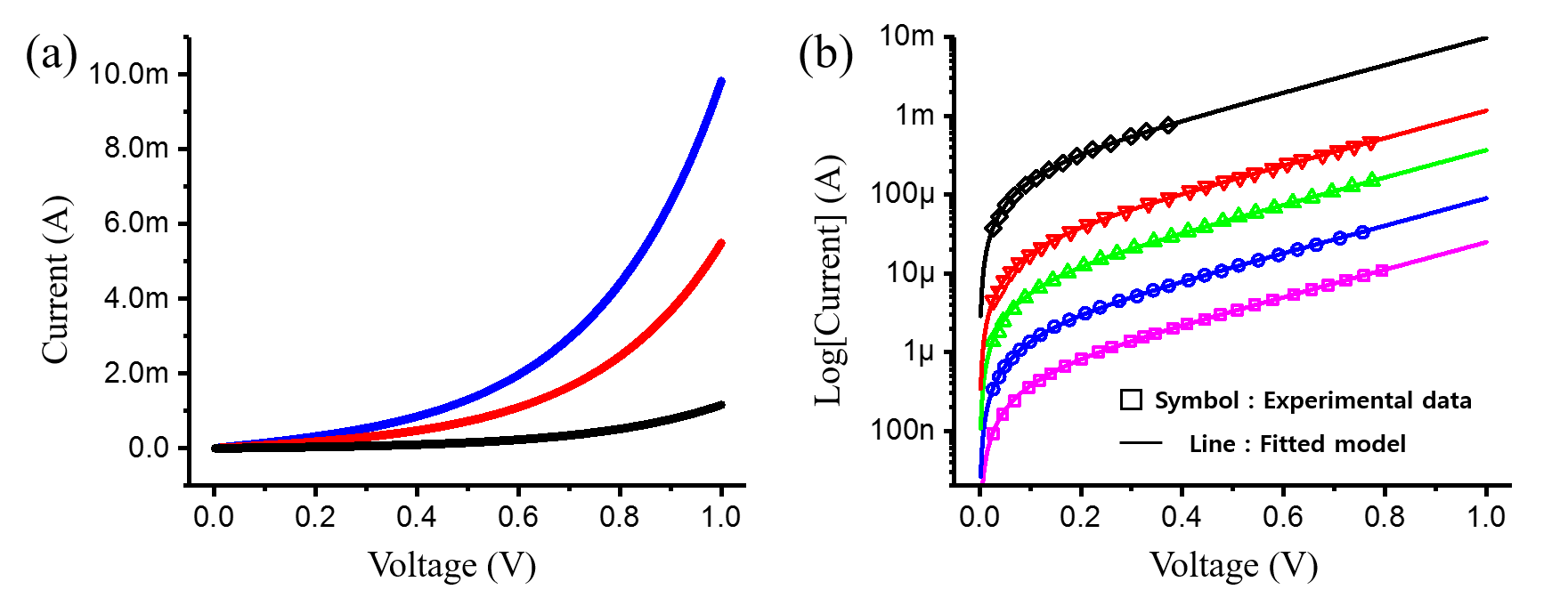}
  \centering
  \caption{\small (a) I-V curve of a real RRAM device in different resistance states. (b) Experimental data and fitted curve of the device under various resistance states \cite{sinh}.}
  \label{real_IV}
\end{figure}
\subsection{Nonlinear I-V Characteristics of RRAM Devices}

Fig. \ref{real_IV} illustrates the I-V characteristics of an actual metal-oxide RRAM device extracted from \cite{sinh}. Each line shows the I-V curve of the RRAM device given a specific sequence of set-voltage pulses. As shown in Fig. \ref{real_IV}(a), the I-V relationship of the RRAM device has an exponential form. \cite{sinh} suggests that the I-V characteristics can be modeled as a empirical model with a sinh function 
\begin{equation}
I(V) = e^{d/d_0}\text{sinh}(BV) \label{eq1}
\end{equation}
where $d$ is average tunneling gap distance, $d_0$ is a fitting parameter, and $B$ is a constant. Measurement results and the values from the empirical model match well as shown in Fig. \ref{real_IV}(b). Each I-V curve corresponding to a different state can be obtained by appropriately determining a state variable $d$.
%(Fig. \ref{Nonlinearity}(a)).

\begin{figure}[t]
  \includegraphics[width=\textwidth, height=0.38\textwidth]{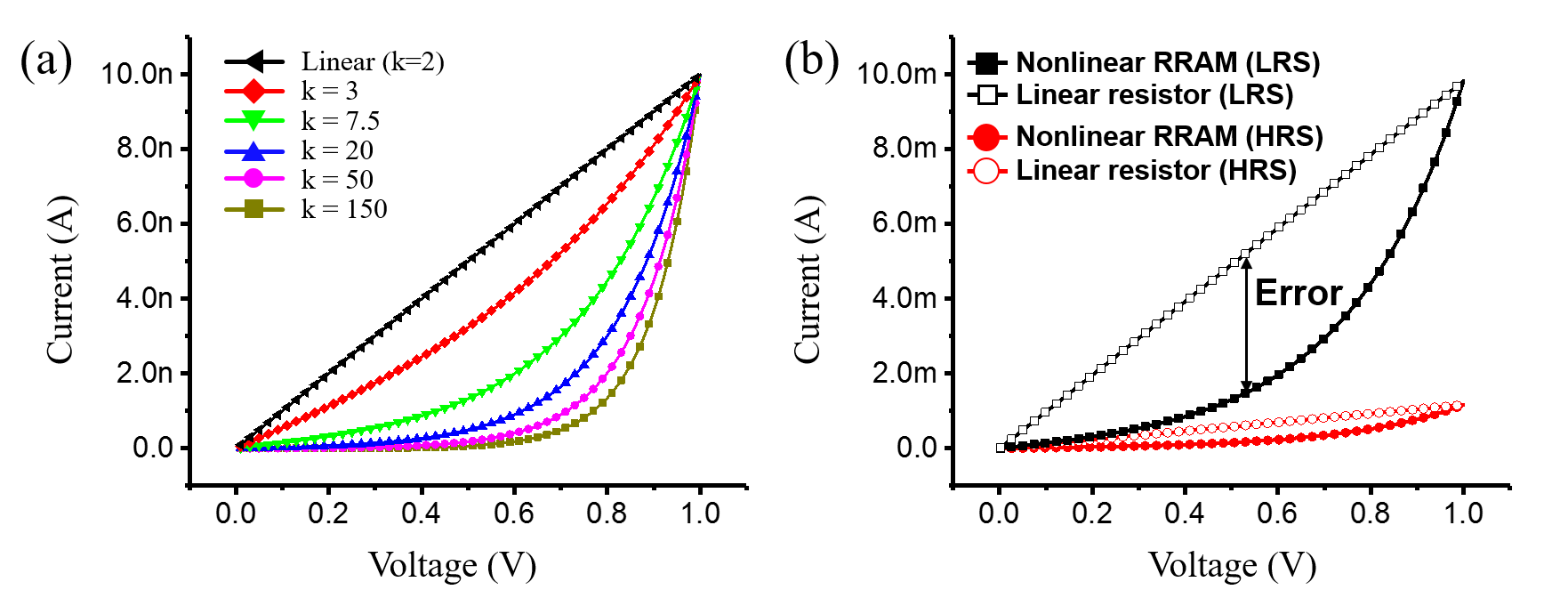}
  \centering
  \caption{\small (a) I-V curves in sinh form with various degrees of nonlinearity. In this case parameter $B$ determines nonlinearity $k$. The model device has $k = 7.5$, leading to $B = 4$ in Eq. (\ref{eq1}). (b) Error between nonlinear I-V curves of RRAM devices and corresponding linear I-V curves of resistors. Hollow symbols stand for the linear resistor model and filled symbols stand for the real device model.}
  \label{Nonlinearity}
\end{figure}

The degree of nonlinearity of an I-V curve can be represented by half-bias nonlinearity \begin{math}k\end{math}, which is defined as 
\begin{equation}
k = \frac{I(V_\text{r,max})}{I({V_\text{r,max}/2)}} \label{eq2}
\end{equation}
where $V_\text{r,max}$ is the maximum value of read voltage that can be used without disturbing the state of the device. A $k$ value does not guarantee unique I-V characteristics of a particular device, as two different devices with the same $k$ value can have different I-V curves. However, it can be still said that $k$ value represents the nonlinearity in some degree. In this paper, we assume \begin{math}V_\text{r,max} = 1\end{math} V for simplicity without loss of generality. Then, \begin{math}k\end{math} can be expressed as a function of $B$ as follows,
\begin{equation}
k = \frac{e^{B}-e^{-B}}{e^{B/2}-e^{-B/2}} \label{eq3}
\end{equation}
When \begin{math}k = 2\end{math}, the device has a linear I-V characteristic. As \begin{math}k\end{math} increases, the I-V characteristic of the corresponding device becomes more nonlinear and exhibits a larger current difference from the linear I-V characteristic with the same resistance state (Fig. \ref{Nonlinearity}(a)). To investigate the \begin{math}k\end{math} values of existing RRAM devices, we surveyed several papers and manually extracted I-V characteristic data of the proposed devices. We could observe that \begin{math}k\end{math} has a wide distribution ranging from ~2.5 \cite{lowk} to ~70 \cite{highk1,highk2}. Among the devices, we chose a device with $k=7.5$ as the model device for the rest of the paper. %Note that I-V nonlinearity is often required to suppress the sneak path problem \cite{lowk2,lowk}

\subsection{Weight Mapping}
ANN utilizes a vector-matrix multiplication of the corresponding input vector and weight matrix to obtain a weighted sum for a layer as
\begin{equation}
\mathbf{s} =  \mathbf{x} \cdot \mathbf{W} \label{eq4}
\end{equation}
with $\mathbf{s}$ as the weighted sum vector, $\mathbf{x}$ as the input vector, and $\mathbf{W}$ as the weight matrix. To implement Eq. (\ref{eq4}) using an emerging NVM crossbar array, each weight in particular row and column of the weight matrix must be mapped to a characteristic parameter of a corresponding device in the crossbar array. In previous approaches \cite{penggu,BSB,DPE} which used RRAM crossbar arrays, weights were mapped to the conductance of the devices assuming linear I-V characteristics as follows,
\begin{equation}
G(w) = \frac{(G_\text{max}-G_\text{min})(w-w_\text{max})}{w_\text{max}-w_\text{min}}+\frac{G_\text{max}(w_\text{max}-w_\text{min})}{G_\text{max}-G_\text{min}} \label{eq5}
\end{equation}
Eq. (\ref{eq5}) takes a \textbf{naive linear-mapping approach} that maps the minimum weight to the minimum conductance and maps the maximum weight to the maximum conductance. In this approach, input vectors can be represented by a set of voltages applied to rows of the crossbar array and the result of the vector-matrix multiplication can be obtained by sampling the current in each column. Previous works mostly relied on the naive mapping and attempted to mimic linear I-V characteristics by limiting the reading voltage into a small linear range \cite{penggu,WRmemristor}. %However, such limitation induced the loss of input bits because of the limited resolution of practical DACs.

\section{Error analysis}

In section 3, we analyze the inference error of the neural networks naively mapped to a RRAM crossbar array consisting of the devices with nonlinear I-V characteristics. To analyze the errors that occur while using full range of $V_\text{read}$, we simulated the consequences of applying the naive mapping method to the devices with the I-V characteristics given in \cite{sinh} (Fig. \ref{real_IV}). Based on the range of device current at 1 V, weights were linearly converted to conductance using Eq. (\ref{eq5}). The empirical I-V model given in Eq. (\ref{eq1}) was used for inference simulations. Fig. \ref{Nonlinearity}(b) shows the mapping strategy.

\begin{figure}[t]
  \includegraphics[width=\textwidth, height=0.437\textwidth]{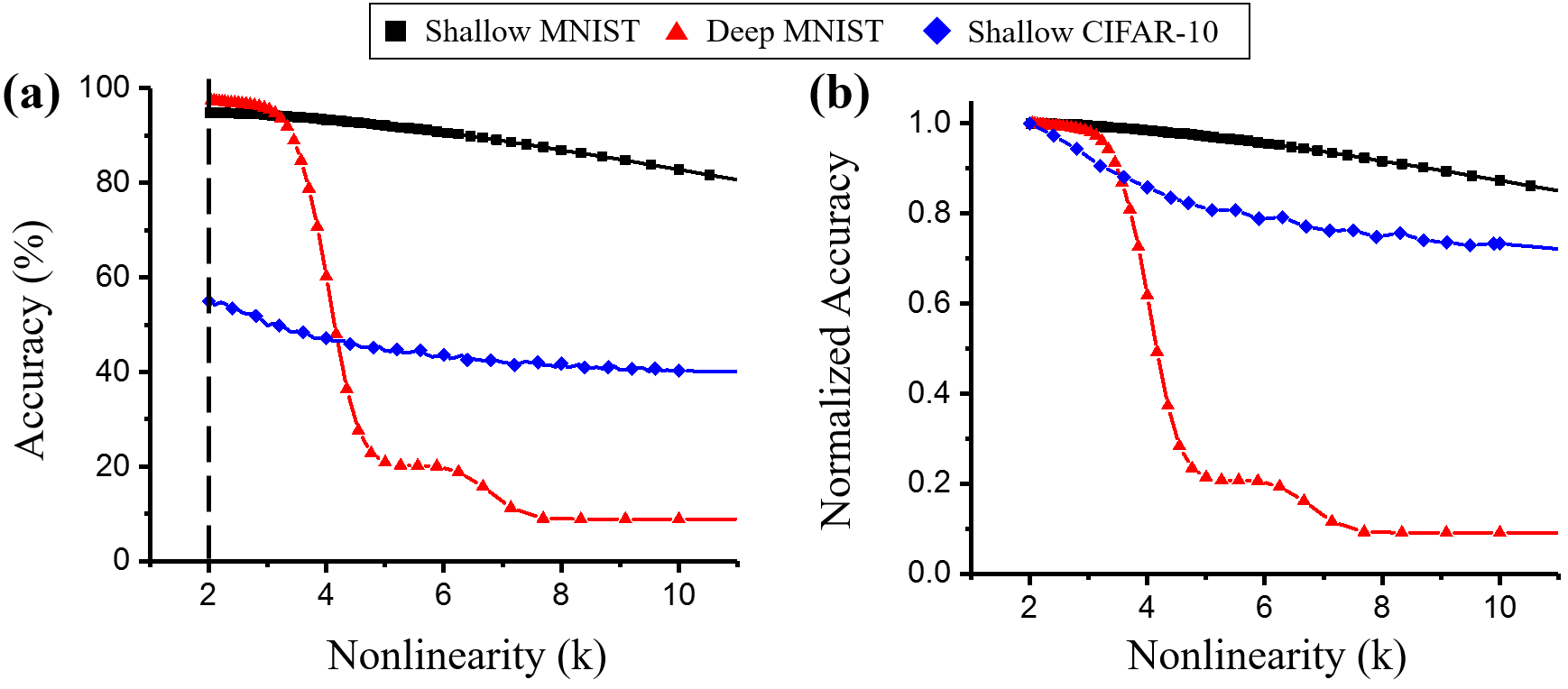}
  \centering
  \caption{\small (a) Accuracy vs. nonlinearity ($k$) and (b) normalized accuracy loss vs. nonlinearity ($k$) for shallow MNIST case (black square), deep MNIST case (red uptriangle), and shallow CIFAR-10 case (blue diamond). (b) describes the accuracy loss relative to the baseline (linear case) accuracy.}
  \label{K_variation}
\end{figure}
\begin{figure}[t]
  \includegraphics[width=0.9\textwidth, height=0.516\textwidth]{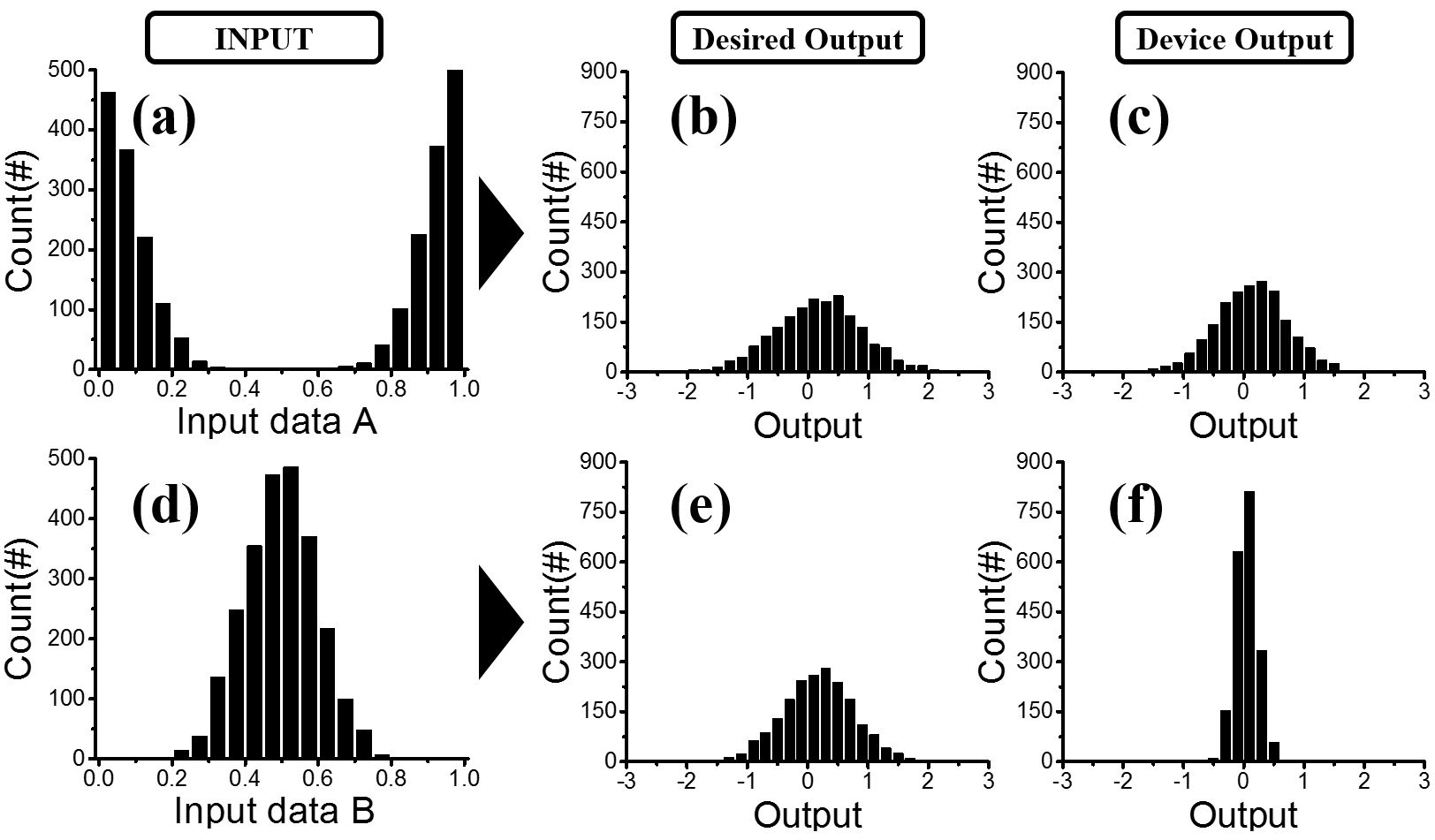}
  \centering
  \caption{\small Distribution of input data A (a) and B (d). Desired output and actual device output for input data A (b,c) and B (e,f).}
  \label{hist6}
\end{figure}
To acquire model weights for analysis, two different Multilayer Perceptron (MLP) networks were trained with MNIST dataset. A shallow network (784-500-250-10, \textbf{shallow MNIST case}) with two hidden layers and a deep network (784-2500-2000-1500-1000-500-10, \textbf{deep MNIST case}) with five hidden layers were each trained with Stochastic Gradient Descent (SGD) using MATLAB. Also, a shallow MLP (2352-4000-1000-4000-10, \textbf{shallow CIFAR-10 case}) was trained with the same method to classify CIFAR-10 dataset for additional analysis.

For each neural network model, the inference accuracy was evaluated using the naive mapping discussed above. Several simulations were performed by sweeping the $k$ values to investigate the relationship between the degree of the inference accuracy degradation and the nonlinearity of the I-V characteristics. For each k, corresponding parameter $B$ in the numerical I-V model Eq. (\ref{eq1}) could be retrieved using Eq. (\ref{eq3}). Evaluation results are illustrated in Fig. \ref{K_variation}. We could observe that overall inference accuracies decrease as the I-V characteristics of the devices become more nonlinear. In addition, the inference accuracy of the deep MNIST case began to drop at lower $k$ compared to the shallow MNIST case. Another observation was that the inference accuracy of the shallow CIFAR-10 case also started to fall at low $k$.

To analyze the cause of accuracy degradation, we investigated the relationship between input value distribution and computation error. Because the current difference between linear and nonlinear I-V curves is the largest at an input voltage of about 0.5 V (Fig. \ref{Nonlinearity}(b)), we speculated that input values around 0.5 V are prone to errors. 

To verify this claim, two distinct distributions were given as input vectors to the second layer of the deep MNIST case. Input data A (Fig. \ref{hist6}(a)) was generated by truncating a normal distribution of numbers so that all data were around 0 and 1. Input data B (Fig. \ref{hist6}(d)) was generated as a normal distribution of numbers with a mean of 0.5 to let all data be around 0.5. The result showed that when input data A was fed to the layer, an output distribution similar to that of the ideal vector-matrix multiplication appeared (Fig. \ref{hist6}(b), (c)). While, feeding input data B to the layer resulted in an output distribution with large error (Fig. \ref{hist6}(e), (f)). Based on this observation, we investigated two representative cases that can induce mid-range input and cause inference accuracy degradation.
\begin{figure}[t]
  \includegraphics[width=\textwidth, height=0.435\textwidth]{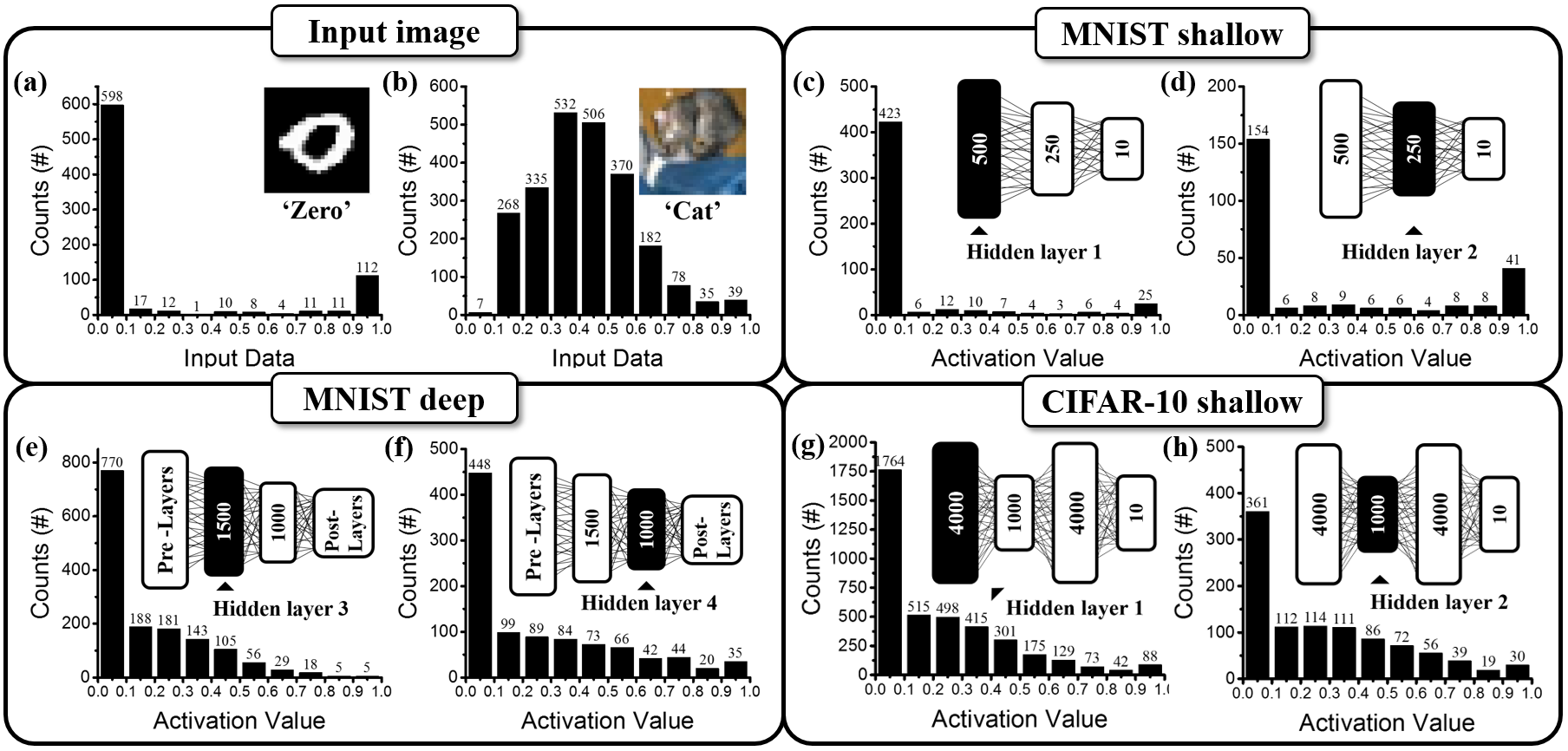}
  \centering
  \caption{\small Distributions of input data in (a) MNIST dataset and (b) CIFAR-10. Activation values of (c) first hidden layer and (d) second hidden layer of shallow MNIST network, (e) third hidden layer and (f) fourth hidden layer of deep MNIST network, and (g) first hidden layer and (h) second hidden layer of shallow CIFAR-10 network.}
  \label{hist8}
\end{figure}

\subsection{Impact of Network Depth}
In the shallow MNIST case, the activation values of neurons tend to yield extreme values such as 0 and $V_\text{read}$ (Fig. \ref{hist8}(a), (c), (d)), since the network can easily fall into saturation due to the lack of training parameters. In contrast, a deeper network with an increased number of training parameters can reduce the probability of saturation and induce mid-range activation values. Increased depth of a layer can make the accuracy even worse because computation error due to mid-range activation values can accumulate on several layers. Fig. \ref{hist8}(e), (f) demonstrates that the activation values of the third and fourth hidden layer of the deep MNIST case have relatively more mid-range values compared to shallow networks. This explains why the shallow MNIST case could maintain relatively high accuracy while the deep MNIST case was vulnerable to the I-V nonlinearity, resulting in greater accuracy loss.

\subsection{Impact of Input Data Distribution}
Fig. \ref{hist8}(a), (b) present the distribution of randomly selected training data from the MNIST dataset and the CIFAR-10 dataset. As the CIFAR-10 dataset consists of natural images with various RGB data, it has more mid-range values compared to the MNIST dataset. Such an input data distribution causes degradation in inference accuracy due to imprecise activation values at the first layers of the network. Besides, hidden layers of the ANNs for the CIFAR-10 dataset tend to generate mid-range values as the networks must extract complex features to classify complex images. As a result, the neuron activation values for CIFAR-10 neural networks have large portion of mid-range values even in the shallow network case (Fig. \ref{hist8}(g), (h)).

\section{Proposed Methodology}
Previous approaches attempted to exploit the linear sub-range in the nonlinear I-V curve for more accurate vector-matrix multiplications. However, its effectiveness was only explored for particular empirical I-V models. DAC resolution also arose as a critical problem for the cases with increased nonlinearity. Thus, there is a pressing need to address the I-V nonlinearity without such limitations. Here, we take a totally reverse approach to solve the problem; we suggest to reconstruct the neural network model itself to reflect the I-V nonlinearity by replacing linear vector-matrix multiplications.

Section 4 proposes a method to build an optimized neural network based on a device model. For the optimized neural network, the basic computation block for vector-matrix multiplications of a neural network is replaced by the nonlinear I-V model of a given device. Any device with nonlinear I-V characteristics can be adopted as far as the numerical model of the I-V curve is differentiable. By reconstructing the neural network considering the nonlinear I-V model of a given device, we aim at overcoming the causes of accuracy loss discussed in Section 3 without limiting the functionality of crossbar arrays.

\subsection{Network Construction}

A perceptron produces its activation value using transfer function and activation function as
\begin{equation}
y = g(f(\mathbf{w},\mathbf{x})) \label{eq6}
\end{equation}
with $\mathbf{x}$ as a vector of input values, $\mathbf{w}$ as a vector of weights, function $f$ as the transfer function and function $g$ as the activation function. Conventional ANN uses weighted sum as the transfer function,
\begin{equation}
f(\mathbf{w},\mathbf{x}) = \sum\limits_{i}{w_{i}x_{i}} \label{eq7}
\end{equation}
For the activation function, there are several choices such as sigmoid, tanh, and ReLU. In case of a fully connected layer with $k$ perceptrons for a output vector $\mathbf{y}$ and $i$ inputs, input vector $\mathbf{x}$ is fed to each $k$ perceptrons with corresponding weight vectors $\mathbf{w}_{k}$. Since the weight vector and the input vector are independent, this computation can be simplified to a vector-matrix multiplication by concatenating the weight vectors into a vector matrix $\mathbf{W}$,
\begin{equation}
\mathbf{y} = g(\mathbf{x}\cdot\mathbf{W}) \label{eq8}
\end{equation}

\begin{figure}[t]
  \includegraphics[width=10cm, height=4.626cm]{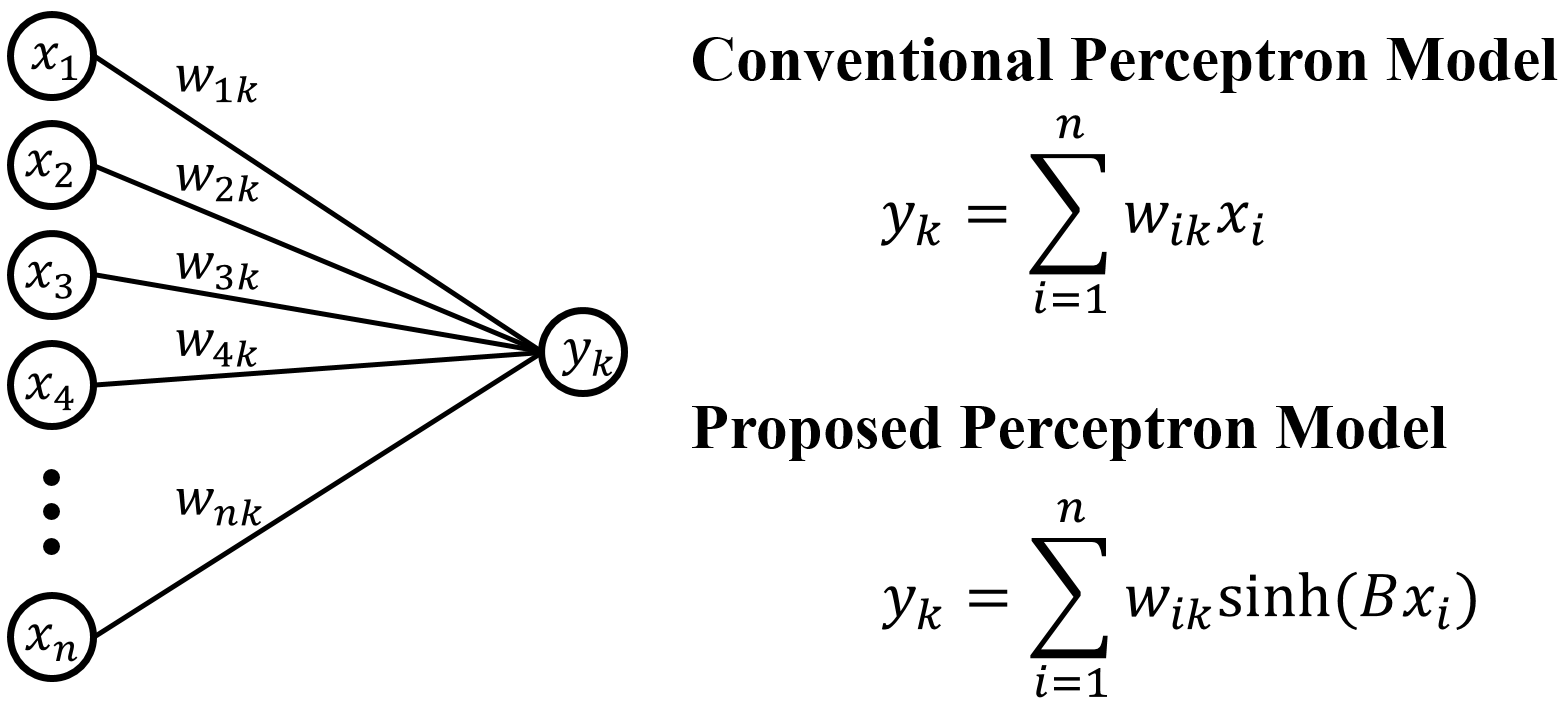}
  \centering
  \caption{\small Conventional and proposed perceptron model}
  \label{perceptron}
\end{figure}

Different from conventional ANN which uses the weighted sum as the transfer function, we propose to use the numerical model of the nonlinear I-V characteristics of a given device as the transfer function of a perceptron. By introducing nonlinearity to the perceptron model itself, we can reduce the gap between the neural network model and the nonlinear I-V characteristics. As an example case, let us construct a neural network using the device from \cite{sinh}.
The empirical model of the device is given as Eq. (\ref{eq1}). The equation can be simplified to
\begin{equation}
I(G,V) = G\text{sinh}(BV) \label{eq9} %\label{eq6}
\end{equation}
As \begin{math}B\end{math} is a characteristic constant of each RRAM device, there are two variables \begin{math}G\end{math} and \begin{math}V\end{math} which determine the output current. Based on the observation, we can define a transfer function as 
\begin{equation}
f(\mathbf{w},\mathbf{x}) = \sum\limits_{i}{w_{i}\text{sinh}(Bx_{i})} \label{eq10} %\label{eqqqqq}
\end{equation}
using conductance $G_{i}$ as weight $w_{i}$ (Fig. \ref{perceptron}). Because the weight vector and the input vector are independent similar to the transfer function of the conventional ANN, we can also concatenate the weight vectors to simplify the computation of an output layer into a vector-matrix multiplication as
\begin{equation}
\mathbf{y} = g(\text{sinh}(B\mathbf{x})\cdot\mathbf{W}) \label{eq11} %\label{eq7}
\end{equation}

Because conductance value is always positive, two adjacent columns of the crossbar array are used to express a single column of weights. We decompose a single weight to a pair of positive and negative sub-weights as
\begin{equation}
w_{i} = w^{+}_{i}-w^{-}_{i} \label{eq12} %\label{eqqwea}
\end{equation}
with $w^{+}_{i}$ as the positive sub-weight and $w^{-}_{i}$ as the negative sub-weight. With this expression, expected computation result can be obtained by subtracting the computation results from two adjacent columns. The transfer function is given as
\begin{equation}
\mathbf{y} = g(\text{sinh}(B\mathbf{x})\cdot\mathbf{W}^{+}-\text{sinh}(B\mathbf{x})\cdot\mathbf{W}^{-}) \label{eq13} %\label{eqqwqw3}
\end{equation}

Besides the modifications, the proposed network use same activation functions, cost functions, optimizers and other components as conventional ANN. For the rest of the paper, we used modified Rectified Linear Unit (ReLU) as the activation function and logistic regression with softmax as the cost function. ReLU function was modified to have a upper bound as the value of maximum read voltage. We used 1 as the upper bound since \begin{math}V_\text{read} = 1\end{math} V. Under the condition, logical output value of a layer could be directly fed into the next layer as the input voltage.

\subsection{Training}
The proposed network can be trained using gradient descent similar to conventional ANN. Gradient for $k_\text{th}$ weight matrix in a $n$-layer network can be derived using chain rule as follows,
\begin{equation}
\frac{dE}{d\mathbf{W}_{k}}=\frac{dE}{d\mathbf{s}_{n}} \cdot \frac{d\mathbf{s}_{n}}{d\mathbf{y}_{n-1}} \cdot \frac{d\mathbf{y}_{n-1}}{d\mathbf{s}_{n-1}} \cdot \frac{d\mathbf{s}_{n-1}}{d\mathbf{y}_{n-2}} \cdot \dots \cdot \frac{d\mathbf{y}_{k+1}}{d\mathbf{s}_{k+1}} \cdot \frac{d\mathbf{s}_{k+1}}{d\mathbf{W}_{k}} \label{eq14} %\label{eq12}
\end{equation}
where $\mathbf{s}$ stands for the result of the transfer function, $\mathbf{y}$ stands for the result of the activation function, and $E$ means the error according to the cost function used. After evaluating the gradient of each weight matrix, gradient descent can be applied to the proposed network to update the weights with $\mu$ as the learning rate: 
\begin{equation}
\mathbf{W}_k^* = \mathbf{W}_k - \mu \frac{dE}{d\mathbf{W}_k} \label{eq15} %\label{eq13}
\end{equation}

Each term in Eq. (\ref{eq14}) can vary depending on the device I-V model used in the neural network. Let us derive the terms for the example case discussed above.
\begin{equation}
\frac{dE}{d\mathbf{s}_n} = (O_\text{true} - O_\text{predict} ) \label{eq16} %\label{eq8}
\end{equation}
The derivative term of error is given as the difference between desired output and predicted output since the example case uses the conventional cross-entrophy loss with softmax function.
\begin{equation}
\frac{d\mathbf{y}_n}{d\mathbf{s}_n} = 
    \begin{cases}
    1, \quad 0\leq \mathbf{s}_n\leq 1\\
    0, \quad else\\
    \end{cases} \label{eq17} %\label{eq11}
\end{equation}
Eq. (\ref{eq17}) shows the derivative term of the activation function used in the example case. At $\mathbf{s}_{n}=0$ and $\mathbf{s}_{n}=1$, we assign 1 to the term for computation although the derivative cannot be explicitly defined.
\begin{equation}
\frac{d\mathbf{s}_n}{d\mathbf{y}_{n-1}} = B\mathbf{W}_{n-1}\cdot \text{cosh}(B\mathbf{y}_{n-1}) \label{eq18}
\end{equation}
Eq. (\ref{eq18}) is derived by taking the derivative of the transfer function with respect to the input vector.
\begin{equation}
\frac{d\mathbf{s}_n}{d\mathbf{W}_{n-1}} = \text{sinh}(B\mathbf{y}_{n-1}) \label{eq19} %\label{eq9}
\end{equation}
Eq. (\ref{eq19}) can be obtained by taking the derivative of the transfer function with respect to the weight matrix. For both Eq. (\ref{eq18}) and Eq. (\ref{eq19}), Eq. (\ref{eq12}) is not considered to simplify the equation. However, simplified sub-weight decomposition can still be used by defining two sub-weights in adjacent columns as
\pagebreak
\begin{equation}
\label{Aplus}
w^+ = 
    \begin{cases}
    w, \quad w\geq 0\\
    0, \quad w<0\\
    \end{cases}
\end{equation}
\begin{equation}
\label{Aminus}
w^- = 
    \begin{cases}
    0, \quad w\geq 0\\
    |w|, \quad w<0\\
    \end{cases} 
\end{equation}
given a weight $w$. With all the derivative terms above, the proposed network can be trained with gradient descent algorithm.

\section{Evaluation}
Several evaluation networks based on the example case in Section 4 were trained and simulated using MATLAB. A shallow (784-500-250-10) network and a deep (784-2500-2000-1500-1000-500-10) network were trained with MNIST dataset. A shallow (2352-4000-1000-4000-10) network for CIFAR-10 dataset was trained for additional analysis.

Because \begin{math}e^{d/d_0}\end{math} in Eq. (\ref{eq1}) is simplified to $G$ and used as the weight in the proposed network, trained weights must be mapped back into the range of the term \begin{math}e^{d/d_0}\end{math} that actual device exhibits. Manual fitting described in Fig. \ref{real_IV} showed that maximum value for \begin{math}e^{d/d_0}\end{math} is \begin{math}e^{-8}\end{math} and the minimum value is \begin{math}e^{-14}\end{math}. \begin{math}k\end{math} value was also measured as 7.5 through fitting.

Sub-weights were mapped to the available range via linear transformation. The linear transformation was as follows,
\begin{equation}
\label{eval_lt}
e^{d/d_{0}}_{\text{mapped}} = W^\pm*\frac{e^{-8}-e^{-14}}{\text{max}(W^{+},W^{-})}+e^{-14}
\end{equation}
Then, simulated current was sampled from each sub-weight column. After subtracting the sampled value of each negative sub-weight column from the sampled value of the corresponding positive sub-weight column, output of the transfer function was retrieved with inverse function of Eq. (\ref{eval_lt}).

\subsection{MNIST Inference Accuracy}
\begin{figure}[ht]
  \includegraphics[width=10.18cm, height=8cm]{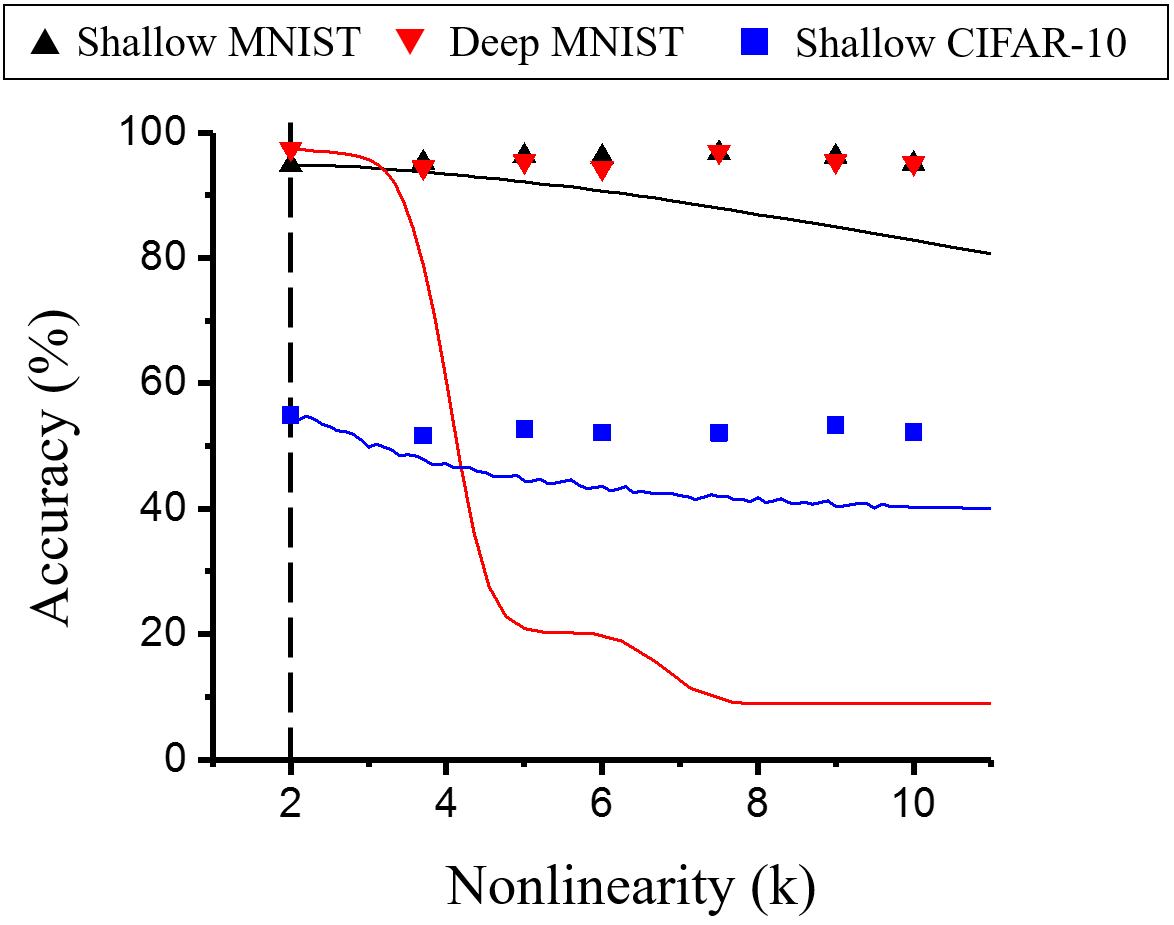}
  \centering
  \caption{\small Inference accuracies of networks trained using the proposed method for various $k$ values. Simulation results from naive mapping are in solid lines and that from proposed mapping are in symbols. Data for Shallow MNIST, Deep MNIST, and Shallow CIFAR-10 cases are in black, red, blue color respectively. Proposed networks (symbol) do not show significant dependency on $k$ value while naively mapped networks (line) do.}
  \label{result}
\end{figure}

Example MLP models were trained with SGD. For the shallow network, we used 30 epochs for training. Learning rate was set to $5\times 10^{-6}$ at first and to $1\times 10^{-6}$ after 16 epochs. The deep network was trained for 65 epochs. Learning rate for the deep network was set to $2\times 10^{-6}$ at first and $7\times 10^{-7}$ after 15 epochs. To demonstrate the robustness against nonlinear I-V characteristics of various devices, model networks with various $k$ values were also tested. Same SGD was used but learning rate varied for each case. Fig. \ref{result} demonstrates the inference evaluation results. 

The proposed network did not show noticeable degradation in accuracy for various $k$ values while networks based on naive mapping exhibited drastic accuracy loss as $k$ increased. Accuracies for the case with the example device ($k=7.5$) were as shown in Table \ref{table1}. This result shows that the proposed network can minimize the error demonstrated in Section 3.1.

\begin{table}[b]
\centering
\caption{\small Evaluation results for device with $k=7.5$ (\%)}
\begin{tabular}{|c|c|c|c|} \hline
 &Shallow MNIST & Deep MNIST & Shallow CIFAR-10\\ \hline
 Ideal case & 94.8&97.43&54.97 \\ \hline
Naive mapping & 87.90&9.05&41.94\\ \hline
Proposed network & 96.74&96.91&52.09\\ \hline
\end{tabular} \label{table1}
\end{table}
\subsection{CIFAR-10 Accuracy}
Another MLP with the proposed model was trained for CIFAR-10 dataset as a proof-of-concept model. The training set was randomly cropped into a set of 28x28 images and exposed to random image distortions. The random distortions included horizontal flips and contrast and brightness adjustments. Then, the images were divided by 255 to ensure input range between 0 and 1 because the raw data are unsigned 8-bit integers. The MLP was trained for this image dataset using SGD for 100 epochs. Learning rate was fixed to $1\times 10^{-6}$. MLP networks with varying $k$ values for CIFAR-10 dataset were also trained.

CIFAR-10 classification results also showed that the accuracy of the proposed network does not vary much over wide range of the $k$ value. As illustrated in Table \ref{table1}, the proposed scheme achieved better inference accuracy than the naive mapping case. Since MLP is not capable of achieving high inference accuracy for such complex task, inference accuracies of both ideal and proposed network were limited. Still, the result demonstrates that the proposed network can also circumvent the error discussed in Section 3.2.

%\subsection{Sensitivity to Resistance Variation}
%We also investigated how much device variations can degrade the inference accuracies of neural networks. All of the networks trained using proposed method and previous approaches were tested with random noises. Since precise control of RRAM conductance is difficult, device variations can occur. To mimic device variations, a matrix of Gaussian random numbers was multiplied elementwise to obtained \begin{math}e^{d/d_0}\end{math} value matrix. Several simulations were conducted with varying Gaussian random distributions which have 1 as mean and sigma from 0 to 0.25. For all the cases of the proposed network, no significant accuracy loss (\begin{math}<\sim\end{math}1\%) was observed. Linear MLPs were simulated with imaginary linear devices to compare the trends of accuracy degradation as device variation increases (Fig. \ref{device_variation}).%

%\begin{figure}[t]
  %\includegraphics[width=6cm, height=4.9cm]{device_variation}
  %\centering
  %\vspace{-3mm}
  %\caption{\small Inference accuracy trend according to increasing device variation. Cross %points depict the data from baseline MLP and lines show the results from proposed networks.}
 % \vspace{-5mm}
 % \label{device_variation}
%\end{figure}

\section{Application to more complex I-V model}

In addition to the device model derived in \cite{sinh}, there are several other resistive devices and corresponding empirical I-V models \cite{selector,PCRAM}. While each device has a different I-V model, the proposed method is generally applicable as far as the I-V model is differentiable.

\begin{figure}[t]
  \includegraphics[width=\textwidth, height=0.38\textwidth]{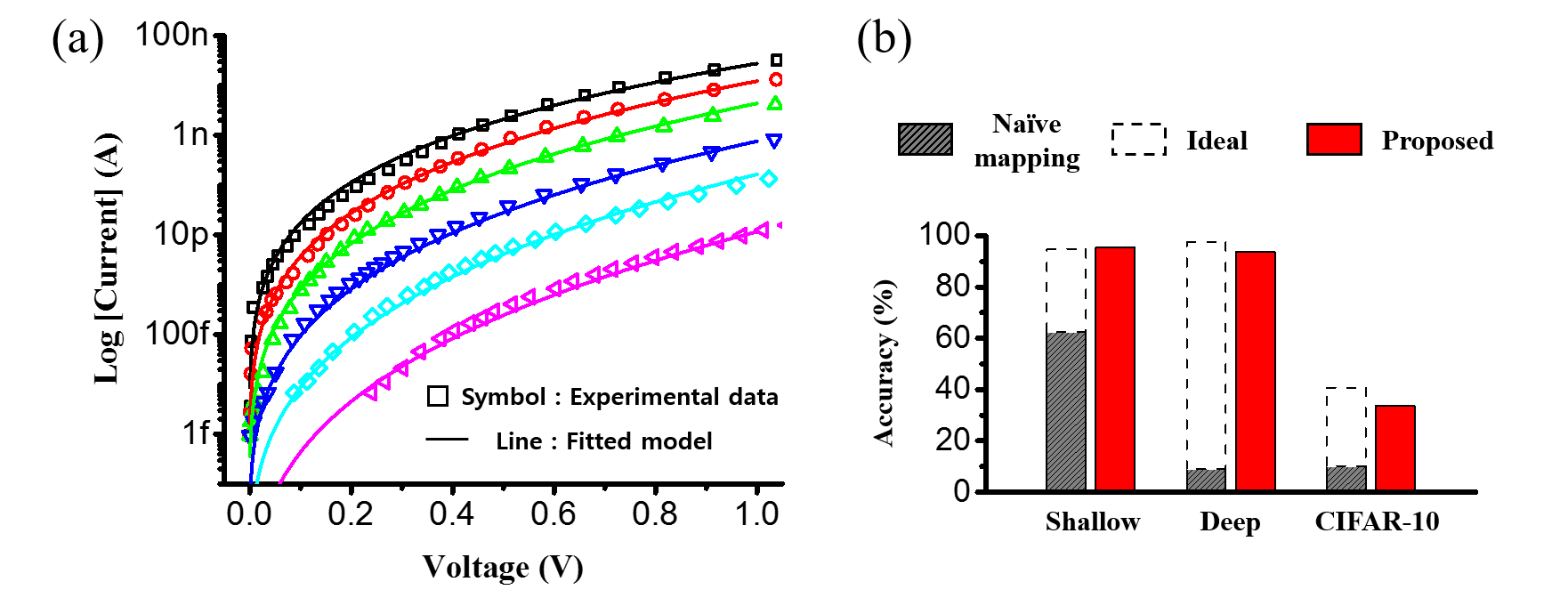}
  \centering
  \caption{\small (a) Experimental data and fitted curve of another device under various resistance states \cite{hwang}. (b) Accuracy comparison between previous and proposed approaches. This result shows that our proposed method can also be applied to complex device models.}
  \label{hwang_fig}
\end{figure}

To demonstrate the validity of this claim, we built and tested three MLPs using the device model obtained by manually fitting the I-V characteristics of the device in \cite{hwang} as the transfer function. We intentionally made a complex empirical model to verify our assertion. The empirical model is as follows,
\begin{equation}
I(w,V) = e^{Aw+B}(e^{CV^{w+D}}-1) \label{eq23}
\end{equation}
with $A = -53.59$, $B = -37.058$, $C = 20$, $D = 0.2$, and $0<W<0.15$. Note that the model is more complex than Eq. (\ref{eq1}). Fitting result of the model was as shown in Fig. \ref{hwang_fig}(a). MLPs for MNIST and CIFAR-10 with same structures as the ones described earlier were built and evaluated. To train the network using gradient descent, same chain rule as Eq. (\ref{eq14}) was derived. Since the cost function and activation function were the same, derivative terms Eq. (\ref{eq16}) and Eq. (\ref{eq17}) remained the same. However, for other terms, the derivation process was very different because the weight term and the input voltage term were correlated. Because of the correlation, we had to consider the simplified sub-weight decomposition while choosing the transfer function as 
%\begin{equation}
%\label{eq24}
%y_{j} = 
    %\begin{cases}
    %g\big(\sum\limits_{i} (e^{Aw_{i,j}+B}(e^{Cx_{i}^{w_{i,j}+D}}-1)-e^{B}(e^{Cx_{i}^{D}}-1))\big), \quad w_{i,j}\geq 0\\
    %g\big(\sum\limits_{i} (e^{B}(e^{Cx_{i}^{D}}-1)-e^{-Aw_{i,j}+B}(e^{Cx_{i}^{-w_{i,j}+D}}-1))\big), \quad w_{i,j}<0\\
%    \end{cases} 
%\end{equation}
\begin{equation}
\label{eq24}
s = I(w^{+},V)-I(w^{-},V)
\end{equation}
where Eq. (\ref{Aplus}) and Eq. (\ref{Aminus}) lead to
\begin{equation}
\label{eq25}
s_{n,j} = 
    \begin{cases}
    \sum\limits_{i} \big(e^{Aw_{n-1,i,j}+B}(e^{Cx_{n-1,i}^{w_{n-1,i,j}+D}}-1)-e^{B}(e^{Cx_{n-1,i}^{D}}-1)\big), \quad w_{n-1,i,j}\geq 0\\
    \sum\limits_{i} \big(-e^{-Aw_{n-1,i,j}+B}(e^{Cx_{n-1,i}^{-w_{n-1,i,j}+D}}-1)+e^{B}(e^{Cx_{n-1,i}^{D}}-1)\big), \quad w_{n-1,i,j}<0\\
    \end{cases} 
\end{equation}
for the $j_{\text{th}}$ neuron in $n_{\text{th}}$ layer. Since the transfer function must be computed element-wise, derivative terms of the transfer function with respect to the input vector and the weight vector had to be also expressed element-wise as
\begin{equation}
\frac{ds_{n,j}}{dw_{n-1,i,j}} = 
    \begin{cases}
    \sum\limits_{i} e^{Cx_{n-1,i}^{w_{n-1,i,j}+D}+Aw_{n-1,i,j}+B}(Cx_{n-1,i}^{w_{n-1,i,j}+D}ln{V}-A), \quad w_{n-1,i,j}\geq 0\\
    \sum\limits_{i} e^{Cx_{n-1,i}^{-w_{n-1,i,j}+D}-Aw_{n-1,i,j}+B}(Cx_{n-1,i}^{-w_{n-1,i,j}+D}ln{V}-A), \quad w_{n-1,i,j}<0\\
    \end{cases}
    \label{eq26}
\end{equation}
\begin{equation}
\label{eq27}
\frac{ds_{n,j}}{dx_{n-1,i}} = 
    \begin{cases}
    \!\begin{aligned}
    \sum\limits_{i} x_{n-1,i}^{w_{n-1,i,j}+D-1}e^{Cx_{n-1,i}^{w_{n-1,i,j}+D}+Aw_{n-1,i,j}+B-1}(Cw_{n-1,i,j}+CD)-CDx_{n-1,i}^{D-1}&e^{Cx_{n-1,i}^{D}+B},\\
    &w_{n-1,i,j}\geq 0\\
    \sum\limits_{i} -x_{n-1,i}^{-w_{n-1,i,j}+D-1}e^{Cx_{n-1,i}^{-w_{n-1,i,j}+D}-Aw_{n-1,i,j}+B-1}(-Cw_{n-1,i,j}+CD)+CD&x_{n-1,i}^{D-1}e^{Cx_{n-1,i}^{D}+B},\\
    &w_{n-1,i,j}<0
    \end{aligned}
    \end{cases}
\end{equation}
Since Eq. (\ref{eq27}) was not defined for $x=0$, we substituted 0 as the term for such cases during the training phase. Using the terms above, we could obtain the accuracy results as shown in Fig. \ref{hwang_fig}(b). The results showed that the proposed method is applicable to very complex nonlinear device I-V model.

\section{Conclusion \& Future Research}
In this paper, we aimed for accurate computation of neural networks using emerging NVM crossbar arrays. We first analyzed the cause of inaccuracy in using naive mapping method. Simulation results showed that mid-range activation values induce computation error for complex tasks and networks. To overcome the accuracy degradation due to the nonlinear I-V characteristics of emerging NVM devices, we proposed a method to construct neural networks optimized to the characteristics. Neural networks based on empirical models of two RRAM devices were trained and tested to classify MNIST and CIFAR-10 dataset using the proposed approach. Results showed that proposed networks could achieve inference accuracies comparable to the baseline. Also, proposed networks did not show accuracy degradation for a variety of nonlinearity values while naive mapping showed significant accuracy loss.

Classification accuracy for the shallow CIFAR-10 case was limited due to the inherent lack of computing power of the MLP structure. Thus, next step of this research will be applying the proposed methodology to more complex neural networks such as CNN and RNN. Also, we did not take into account other non-ideal characteristics of emerging NVM arrays such as I-R drop for simulations. Another next step of this work will be to simulate and analyze the effects of such characteristics.
\vspace{-3mm}
\section{Acknowledgement}

This work was in part supported by the Ministry of Science, ICT and Future Planning of South Korea under the "IT Consilience Creative Program" (IITP-R0346-16-1007) and "Nano-Material Technology Development Program" (No. 2016910249) through the National Research Foundation of Korea (NRF). It is also in part supported by the Industrial Technology Innovation Program (10067764) funded by the Ministry of Trade, Industry \& Energy (MOTIE, Korea).
\vspace{-3mm}
\footnotesize
\bibliography{main}{}
\bibliographystyle{ieeetr}

\end{document}